# “Betweenness Centrality” as an Indicator of the “Interdisciplinarity” of Scientific Journals

Loet Leydesdorff

Amsterdam School of Communications Research (ASCoR), University of Amsterdam

Kloveniersburgwal 48, 1012 CX Amsterdam, The Netherlands

loet@leydesdorff.net; http://www.leydesdorff.net

**Abstract**

In addition to science citation indicators of journals like impact and immediacy, social network analysis provides a set of centrality measures like degree, betweenness, and closeness centrality. These measures are first analyzed for the entire set of 7,379 journals included in the *Journal Citation Reports* of the *Science Citation Index* and the *Social Sciences Citation Index* 2004, and then also in relation to local citation environments which can be considered as proxies of specialties and disciplines. Betweenness centrality is shown to be an indicator of the interdisciplinarity of journals, but only in local citation environments and after normalization because otherwise the influence of degree centrality (size) overshadows the betweenness-centrality measure. The indicator is applied to a variety of citation environments, including policy-relevant ones like biotechnology and nanotechnology. The values of the indicator remain sensitive to the delineations of the set because of the indicator’s local character. Maps showing interdisciplinarity of journals in terms of betweenness centrality can be drawn using information about journal citation environments which is available online.

**Keywords**: centrality, betweenness, interdisciplinarity, journal, citation, indicator

## 1. Introduction

Ever since Garfield (1972; Garfield & Sher, 1963) proposed impact factors as indicators for the quality of journals in evaluation practices, this measure has been heavily debated. Impact factors were designed with the purpose of making evaluation possible (e.g., Linton, 2006). Other indicators (e.g., Price's [1970] immediacy index) were also incorporated into the *Journal Citation Reports* of the *Science Citation Index*, but were coupled less directly to library policies and science policy evaluations (Moed, 2005; Monastersky, 2005; Bensman, forthcoming).

Soon after the introduction of the *Science Citation Index*, it became clear that publication and citation practices are field-dependent (Price, 1970; Carpenter & Narin, 1973; Gilbert, 1977; Narin, 1976). Hirst (1978), therefore, suggested constructing discipline-specific impact factors, but their operationalization in terms of discipline-specific journal sets has remained a problem. Should such sets be defined with reference to the groups of researchers under evaluation (Moed *et al*., 1985) or rather in terms of the aggregated citation patterns among journals (Pinski & Narin, 1976; Garfield, 1998)? How can one disentangle the notion of hierarchy among journals and the juxtaposition of groups of journals in the various disciplines (Leydesdorff, 2006)?

Furthermore, as Price (1965) noted, different types of journal publications within similar fields can be expected to vary also in terms of their citation patterns. Within each field, some journals follow developments at the research front (e.g., in the form of letters), while other journals (e.g., review journals) have a longer-term scope. Thirdly, journals differ in terms of their "interdisciplinarity," with *Nature* and *Science* as the prime examples (Narin *et al*., 1972), while others include sections of both general interest and disciplinary affiliations (e.g., *PNAS* and the *Lancet*). In addition to the "multidisciplinarity" or "interdisciplinarity" of journals at a general level, "interdisciplinarity" can also occur at the very specialized interface between established fields of science, as in the case of biotechnology and nanotechnology.

Three indicators of journals were codified in the ISI databases: impact factors, immediacy indices, and the so-called subject categories. These indicators are based on the *Journal Citation Reports*, which offer aggregated citation data among journals. However, the subject categorization of the ISI has remained the least objective among these indicators because the indicator is not citation-based. The ISI-staff assigns journals to subjects on the basis of a number of criteria, among which are the journal's title, its citation patterns, etc. (McVeigh, *personal communication*, 9 March 2006).

An unambiguous categorization of the journal set in terms of subject matters seems impossible because of the fuzziness of the subsets (Bensman, 2001). In addition to intellectual categories, journals belong to nations, publishing houses, and often to more than a single discipline (Leydesdorff & Bensman, 2006). The potential "interdisciplinarity" of journals makes it difficult to compare journals as units of analysis within a specific reference group of "disciplinary" journals.

"Interdisciplinarity" is often a policy objective, while new developments may take place at the borders of disciplines (Caswill, 2006; Zitt, 2005). New developments may lead to new journal sets or be accommodated within existing ones (Leydesdorff *et al*., 1994). For example, recent developments in nanotechnology have evolved at interfaces among applied physics, chemistry, and the material sciences. The delineation of a journal set in nanotechnology is therefore not a *sine cure*, while in the meantime a much more discrete set of journals in biotechnology has evolved. Existing classifications may have to be revised and innovated from the perspective of hindsight (Leydesdorff, 2002). The U.S. Patent and Trade Office, for example, has launched a project to reclassify its existing database using "nanotechnology" as a new category at the level of individual patents.

Reclassification at the level of individual articles would mean changing the (controlled) keywords with hindsight (Lewison & Cunningham, 1988). However, this is unnecessary since scientific articles are organized into journals by a strong selection process of submission and peer review. The recursive selection processes lead to very strong structures and correspondingly skewed distributions. Garfield (1972, at p. 476) argued that a multidisciplinary core for all of science comprises no more than thousand journals.

The citation structures among journals are updated each year because of changes in citation practices. However, in the case of "interdisciplinary" developments the classification may be more ambiguous because different traditions and standards are interfaced. Cross-links (e.g., citations) provide inroads for change in an otherwise (nearly) decomposable system (Simon, 1973).

The development of a measure of interdisciplinarity at the level of journals derived from this destabilizing effect on citation structures could be extremely useful as an early-warning indicator of new developments. In a previous attempt to develop such indicators, Leydesdorff *et al*. (1994) were able to show that new developments can be traced in terms of deviant being-cited patterns in various groups of neighboring journals. However, the opposite effect, namely that this deviant pattern also indicates new developments, could not be shown (Leydesdorff, 1994; Van den Besselaar & Heimeriks, 2001). Cross-links may have other functions as well. Like most research in the bibliometric field, these analyses of interdisciplinarity were based on the assumption that journals can be grouped either using the ISI subject categories (e.g., Leeuwen & Tijssen, 2000; Morillo *et al.*, 2003) or on the basis of clustering citation matrices (Doreian & Farraro, 1985; Leydesdorff, 1986; Tijssen *et al*., 1987).

Before one can delineate groups of journals in "interdisciplinary" fields, one would need an indicator of "interdisciplinarity" at the level of individual journals. To what extent do articles in a specific journal feed into or draw upon different intellectual traditions? The focus on the position of individual agents in networks—in this case journals—has been developed in social network analysis more than in scientometrics (Otte & Rousseau, 2002).

**2. Centrality Measures in Social Network Analysis**

Social network analysis has developed as a specialty in parallel with scientometrics since the late 1970s. In a ground-laying piece, Freeman (1977) developed a set of measures of centrality based on betweenness. Freeman stated that "betweenness" as a structural property of communication was elaborated in the literature as the first measure of centrality (Bavelas, 1948; Schimbel, 1953). In a follow-up paper, Freeman (1978) gradually elaborated four concepts of centrality in a social network, which have since been further developed (Hanneman & Riddle, 2005; De Nooy *et al*., 2005):

1. centrality in terms of "degrees:" in- and outgoing information flows from each node as a center;
2. centrality in terms of "closeness," that is, the distance of an actor from all other actors in a network. This measure operationalizes the expected reach of a communication;

3. centrality in terms of "betweenness," that is, the extent that the actor is positioned on the shortest path ("geodesic") between other pairs of actors in the network; and
4. centrality in terms of the projection on the first "eigenvector" of the matrix.

These measures and their further elaboration into relevant statistics were conveniently combined in the software package UCINet that Freeman and his collaborators have developed since the 1980s (Bonacich, 1987; Borgatti *et al*., 2002; Otte & Rousseau, 2002). A number of visualization programs for networks like Pajek and Mage interface with UCINet. The visualization and the statistics have become increasingly integrated.

Centrality in terms of degree is easiest to grasp because it is the number of relations a given node maintains. Degree can further be differentiated in terms of "indegree" and "outdegree," that is, incoming or outgoing relations. In the case of a citation matrix, the total number of references provided by a textual unit of analysis (e.g., an article or a journal) can then be considered as its outdegree, and instances of its being cited as the indegree. Degree centrality is often normalized as a percentage of the degrees in a network.

"Betweenness" is a measure of how often a node (vertex) is located on the shortest path (geodesic) between other nodes in the network. It thus measures the degree to which the node under study can function as a point of control in the communication. If a node with a high level of betweenness were to be deleted from a network, the network would fall apart into otherwise coherent clusters. Unlike degree, which is a count, betweenness is normalized by definition as the proportion of all geodesics that include the vertex under study. If $g_{ij}$ is defined as the number of geodesic paths between $i$ and $j$, and $g_{ikj}$ is the number of these geodesics that pass through $k$, $k$'s betweenness centrality is defined as (Farrall, 2005):

$$\sum_i \sum_j \frac{g_{ikj}}{g_{ij}}, \; i \neq j \neq k$$

"Closeness centrality" is also defined as a proportion. First, the distance of a vertex from all other vertices in the network is counted. Normalization is achieved by defining closeness centrality as the number of other vertices divided by this sum (De Nooy *et al.*, 2005, p. 127). Because of this normalization, closeness centrality provides a global measure about the position of a vertex in the network, while betweenness centrality is defined with reference to the local position of a vertex.

Eigenvector analysis brings us back to approaches that are familiar from multivariate analysis. Principal component and factor analysis decompose a matrix in terms of the latent eigenvectors which determine the *positions* of nodes in a network, while graph analysis begins with the vectors of observable *relations* among nodes (Burt, 1982). How can these be grouped bottom-up using algorithms? For example, core-periphery relations can be made visible using graph-analytical techniques, but not by using factor-analytical ones (Wagner & Leydesdorff, 2005).

Betweenness is a relational measure. One can expect that a journal which is "between" will load on different factors because it does not belong to one of the dense groups, but relates them. The factor loadings of such journals may depend heavily on the factor-analytic *model* (e.g., the number of factors to be extracted by the analyst). For example, one might expect inter-factorial complexity among the factor loadings in the case of inter- or multidisciplinary journals (Van den Besselaar & Heimeriks, 2001; Leydesdorff, 2004). Closeness is less dependent on relations between individual vertices because a vertex can be close to two (or more) densily connected clusters. Closeness can thus be expected to provide us with a measure of "multidisciplinarity" within a set while betweenness may provide us with a measure of specific "interdisciplinarity" at interfaces.

## 3. Size, impact, and centrality

While the impact factor and the immediacy index are corrected for size (because the number of publications in the previous two years and the current year, respectively, is used in the denominator; cf. Bensman, forthcoming), centrality measures are sensitive to size. A further complication, therefore, is the possibility of spurious correlations between different centrality measures. Large journals (e.g., *Nature*) which one would expect to be "multidisciplinary" rather than "interdisciplinary," might generate a high betweenness centrality because of their high degree centrality.

Normalization of the matrix for the size of patterns of citation can suppress this effect (Bonacich, *personal communication*, 22 May 2006). Fortunately, there is increasing consensus that normalization in terms of the cosine and using the vector-space model provides the best option in the case of sparse citation matrices (Ahlgren *et al*., 2003;

Chen, 2006; Salton & McGill, 1983). Using the cosine for the visualization, a threshold has to be set because the cosine between citation patterns of locally related journals will almost never be equal to zero. However, the algorithms for computing centrality first dichotomize this matrix.

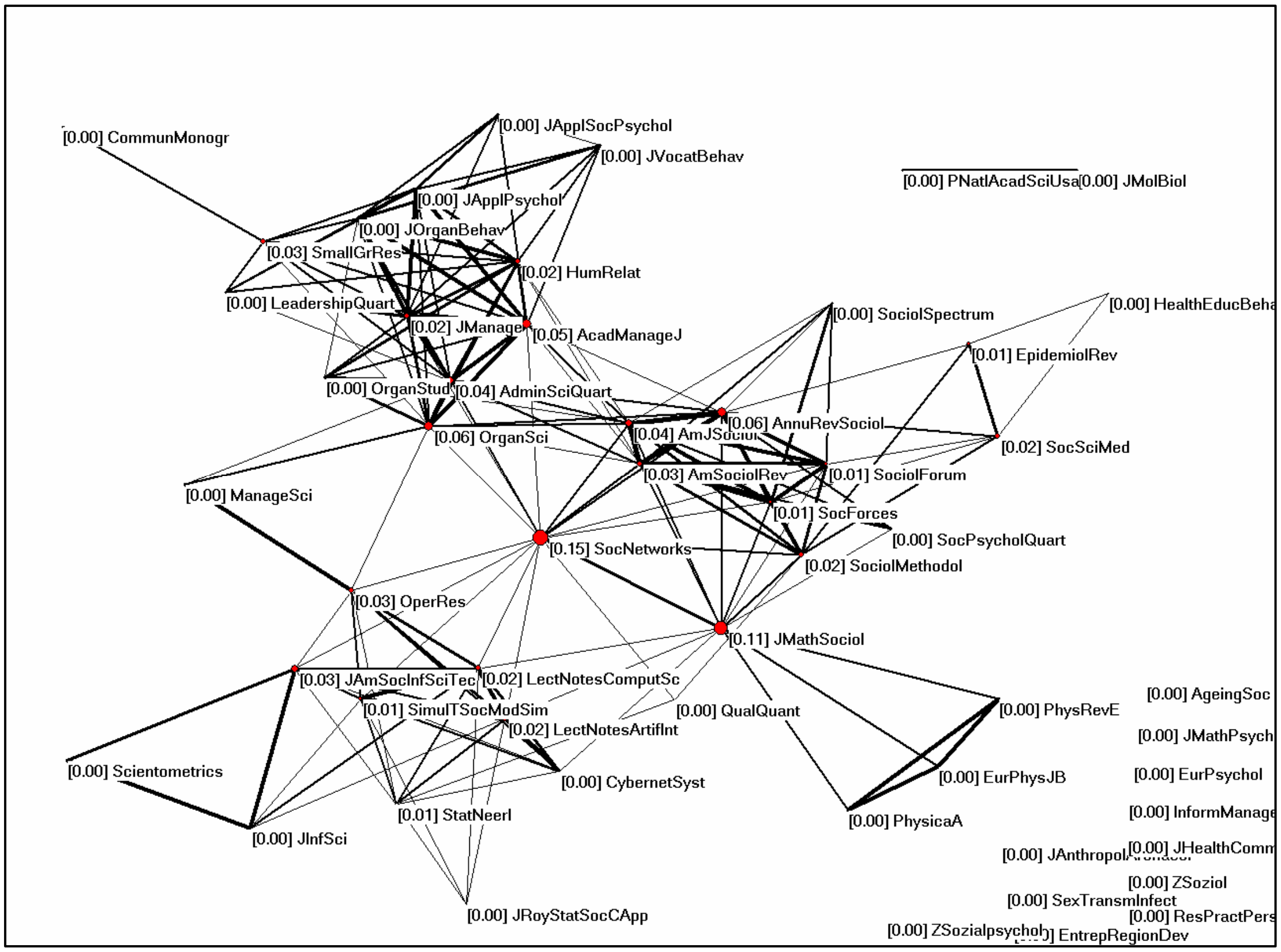

**Figure 1**: Betweenness centrality of 54 journals in the vector space of the citation impact environment of *Social Networks* (cosine ≥ 0.2).

Actually, when I was working with visualizations of cosine-based journal maps (Leydesdorff, forthcoming-a, forthcoming-b), it occurred to me that the interdisciplinarity of journals corresponds with their visible position in the vector space. Figure 1, for example, shows the citation impact environment of *Social Networks* as an example. Among the 54 journals citing *Social Networks* more than once in 2004,[1] this journal is on the shortest path between vertices in 15% of the possible cases, followed by the *Journal of Mathematical Sociology* with a value of 11% on betweenness centrality. The other journals have considerably lower values. The visual pattern of connecting different subgroups also follows the intuitive expectation of "interdisciplinarity" among these journals.

[1] Aggregate values of one are aggregated by the ISI and subsumed under the category of "All others"

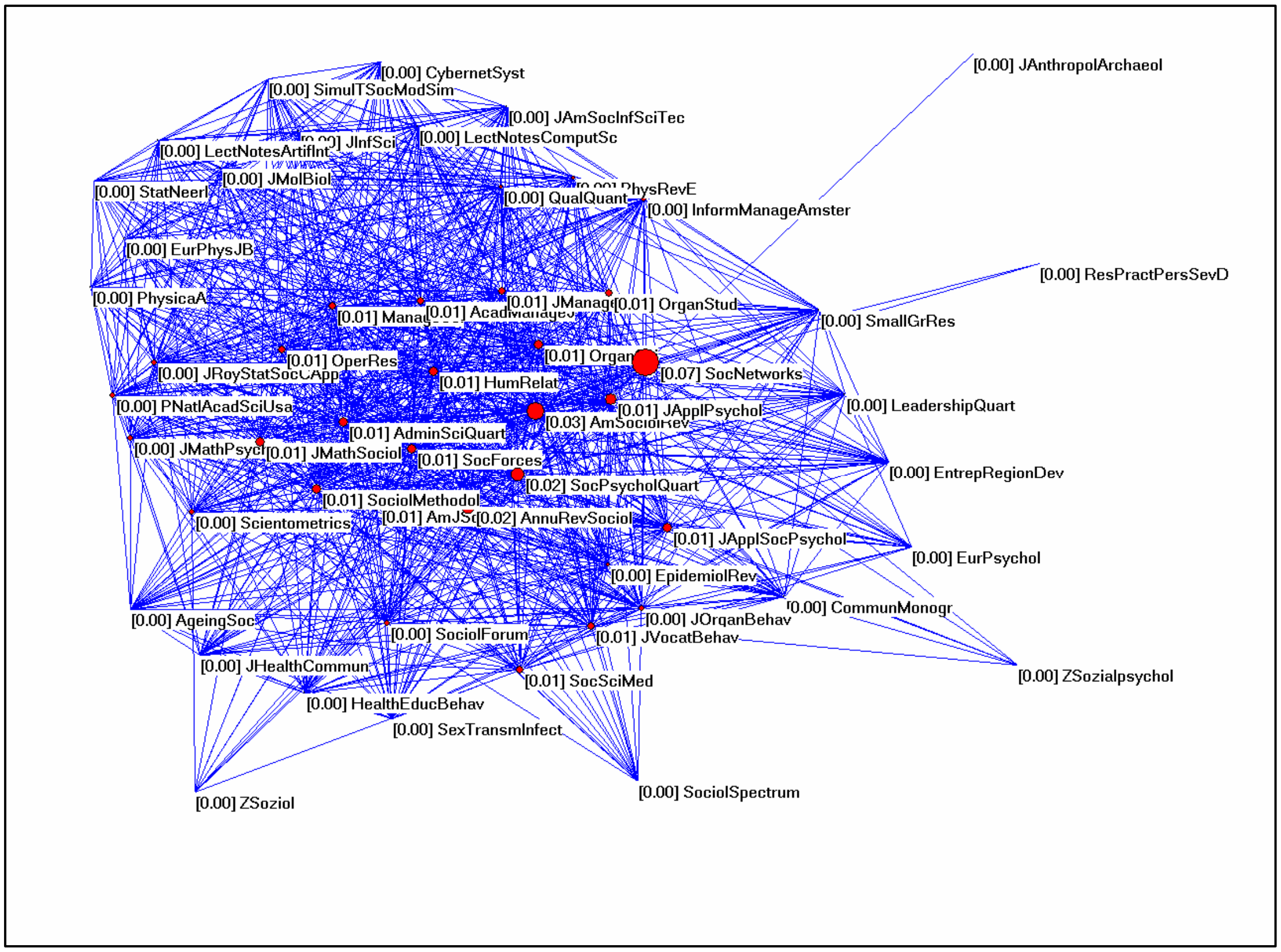

**Figure 2**: Betweenness centrality of *Social Network* in its citation environment before normalization with the cosine.

Figure 2 contrasts this finding with the betweenness centrality in the unnormalized networks. *Social Networks* is still the journal with the largest betweenness value (0.07), but the *Journal of Mathematical Sociology* now has a score of 0.01. This is even lower than the corresponding value for the *American Sociological Review* (0.03). The latter is a much larger journal with a distinct disciplinary affiliation (that is, sociology). In sum, the visualization using unnormalized citation data can be expected to show neither the cluster structure in the data nor betweenness centrality among groups of nodes. One needs a normalization in terms of similarity patterns (using a similarity coefficient like the Pearson correlation or the cosine) to observe the latent structures in this data.

The research question of this paper is to address the phenomenon of betweenness centrality in the vector space systematically. I will first study the different centrality measures in the non-normalized matrix, then in the cosine-normalized one, and finally in a few applications, including some with obvious policy relevance (nanotechnology and biotechnology).

**4. Methods and Materials**

The data was harvested from CD-Rom versions of the *Journal Citation Reports* of the *Science Citation Index* and the *Social Sciences Citation Index* 2004. These two databases cover 5,968 and 1,712 journals, respectively. Since 301 journals are covered by both databases, a citation matrix can be constructed among (5,968 + 1,712 – 301) = 7,379 journals. Seven journals are not processed by the ISI in the "citing" dimension, but we shall focus below on the "cited" dimension of this matrix. This focus enables us to compare the centrality measures directly with well-established science citation indicators like impact factors, immediacy, etc.

Among the 7,379 vectors of the matrix representing the cited "patterns," similarities were calculated using the cosine. Salton's cosine is defined as the cosine of the angle enclosed between two vectors *x* and *y* as follows (Salton & McGill, 1983):

$$\text{Cosine}(x,y) = \frac{\sum_{i=1}^{n} x_i y_i}{\sqrt{\sum_{i=1}^{n} x_i^2}\sqrt{\sum_{i=1}^{n} y_i^2}} = \frac{\sum_{i=1}^{n} x_i y_i}{\sqrt{(\sum_{i=1}^{n} x_i^2)*(\sum_{i=1}^{n} y_i^2)}}$$

The cosine is very similar to the Pearson correlation coefficient, except that the latter measure normalizes the values of the variables with reference to the arithmetic mean (Jones & Furnas, 1987). The cosine normalizes with reference to the geometrical mean. Unlike the Pearson correlation coefficient, the cosine is non-metric and does not presume normality of the distribution (Ahlgren *et al*., 2003). An additional advantage of this measure is its further elaboration into the so-called vector-space model for the visualization (Chen, 2006).

Note that the two matrices—that is, the matrix of citation data and the matrix of cosine values—are very different: the cosine matrix is a symmetrical matrix with unity on the main diagonal, while citation matrices are asymmetrical transaction matrices with usually outliers (within-journal "self"-citations) on the main diagonal (Price, 1981). The topography of the vector space spanned by the cosine values is accordingly different from the topography of the multi-dimensional space spanned by the vectors of citation values.

Subsets can be extracted from the database in order to measure the relations among journals that are citing a specific journal. I shall call these subsets the local citation impact environments of the journal under study. Betweenness centrality and other centrality measures will be different within these local citation environments from their values in the global set because each two journals within a local set can also be related through the mediation of journals outside the subset.

For the computation of centrality measures I use exclusively the methods available within the Pajek environment. This allows for a one-to-one correspondence between the visualizations and the algorithmic results. (The normalizations are sometimes slightly different between UCINet and Pajek.) Although UCINet is faster and richer in providing various computational options, Pajek is currently able to analyze centrality in asymmetrical matrices in both directions. Given our interest in asymmetrical citation matrices, this can be an advantage. The analysis focuses on degree centrality, betweenness centrality, and closeness centrality because eigenvector analysis is used in Pajek only as a means for the visualization. When displaying the citation impact environments (Leydesdorff, forthcoming-a and forthcoming-b), I shall use the vertical size for the relative citation contributions of journals in a specific environment, and the horizontal size for the same measure, but after correction for within-journal citations.

## 5. Centrality at the level of the *Journal Citation Reports*

### 5.1 The asymmetrical citation matrix

The asymmetrical citation matrix contains two structures, one in the "cited" and another in the "citing" dimension of the matrix. Pajek provides options to compute the three centrality measures (degree, betweenness, and closeness) in both directions. Thus, six indicators can be measured across the file. The values on these six indicators can be compared with more traditional science citation indicators like "impact," "immediacy," and "total citations." (The values of the six [two times three] centrality measures for the 7,379 journals are available online at http://www.leydesdorff.net/jcr04/centrality/index.htm .)

**Rotated Component Matrix(a)**

| | Component | | |
|---|---|---|---|
| | 1 | 2 | 3 |
| Number of issues | .924 | | .185 |
| Total number of references (citing) | .909 | .210 | .237 |
| Within journal "self"-citations | .815 | .152 | |
| Betweenness (citing) | .740 | | .103 |
| Total number of citations (cited) | .672 | .639 | |
| Immediacy | | .806 | .267 |
| Impact | | .802 | .295 |
| **Indegree (cited)** | **.405** | **.713** | **.381** |
| **Betweenness (cited)** | **.261** | **.691** | **-.240** |
| **Closeness (cited)** | | | **.776** |
| Closeness (citing) | .190 | .413 | .663 |
| Outdegree (citing) | .498 | .356 | .633 |

Extraction Method: Principal Component Analysis. Rotation Method: Varimax with Kaiser Normalization.
a Rotation converged in 5 iterations.

**Table 1**: Three-factor solution of the matrix of 7,379 journals versus six centrality measures and a number of science (citation) indicators.

Table 1 shows the rotated three-factor solution for the matrix of 7,379 journals versus the various science indicators and centrality measures as variables. Three factors explain 73.5% of the variance. Factor One (46.9%) can be designated as indicating the size of journals, Factor Two (16.4%) registers the effects of citations ("impact," etc.), and Factor Three (10.3%) seems to indicate the reach of a communication through citation. The strong relation between immediacy and impact has previously been noted by Yue *et al.* (2004). The further elaboration of the relation between centrality measures and science citation indicators would lead me beyond the scope of this study.

In Table 1, the three indicators on which we will now focus our attention are shown in boldface. First, one can note the difference in sign for "betweenness centrality" and "closeness centrality" on the third factor, but as expected, this negative correlation is overshadowed by the commonality between "betweenness centrality" and "indegree" on the first two factors.

**Correlations**

| | | Indegree | Betweenness cited | Closeness cited |
|---|---|---|---|---|
| Indegree | Pearson Correlation | 1 | .509(**) | .651(**) |
| | Sig. (2-tailed) | | .000 | .000 |
| | N | 7379 | 7379 | 7379 |
| Betweenness cited | Pearson Correlation | .509(**) | 1 | .210(**) |
| | Sig. (2-tailed) | .000 | | .000 |
| | N | 7379 | 7379 | 7379 |
| Closeness cited | Pearson Correlation | .651(**) | .210(**) | 1 |
| | Sig. (2-tailed) | .000 | .000 | |
| | N | 7379 | 7379 | 7379 |

** Correlation is significant at the 0.01 level (2-tailed).

**Table 2**: Correlations among the centrality measures in the cited dimension ($N$ = 7,379).

Table 2 provides the correlation coefficients among the three centrality measures. Because of the large *N* (= 7,379) all correlations are significant. However, the correlation between closeness and betweenness is considerably lower ($r = 0.21$; $p < 0.01$) than the other correlations ($r > 0.5$; $p < 0.01$).

| | *Indegree* | | *Between-ness* | | *Closeness* |
|---|---|---|---|---|---|
| *Science* | 4904 | *Science* | 0.098921 | *Science* | 0.538172 |
| *Nature* | 4555 | *Nature* | 0.067541 | *Nature* | 0.522138 |
| *P Natl Acad Sci USA* | 3776 | *P Natl Acad Sci USA* | 0.039714 | *P Natl Acad Sci USA* | 0.490666 |
| *Lancet* | 2834 | *Lancet* | 0.013324 | *Lancet* | 0.456274 |
| *New Engl J Med* | 2780 | *JAMA-J Am Med Assoc* | 0.011943 | *New Engl J Med* | 0.453366 |
| *J Biol Chem* | 2674 | *New Engl J Med* | 0.011665 | *JAMA-J Am Med Assoc* | 0.442401 |
| *JAMA-J Am Med Assoc* | 2510 | *Brit Med J* | 0.009516 | *Ann NY Acad Sci* | 0.441714 |
| *Ann NY Acad Sci* | 2375 | *J Am Stat Assoc* | 0.009486 | *J Biol Chem* | 0.440729 |
| *Brit Med J* | 2228 | *Ann NY Acad Sci* | 0.008139 | *Brit Med J* | 0.433717 |
| *Biochem Bioph Res Co* | 2075 | *J Biol Chem* | 0.007159 | *Biochem Bioph Res Co* | 0.420714 |

**Table 3**: Top-10 journals on three network indicators of centrality in the being-cited direction.

Table 3 shows the ten journals with highest values on these three indicators. The set for the "indegree" overlaps completely with "closeness," and these two sets differ only by a single journal from the list for "betweenness:" the *Journal of the American Statistical Association* is included in the latter set, while *Biochemical and Biophysical Research Communications* is not included in this list. In other words, the three measures may indicate different dimensions, but they do not discriminate sufficiently among one another to provide us with a measure of "interdisciplinarity" or "multidisciplinarity" at the level of the file.

## 5.2 The centrality measures in the vector space

Let us turn now to the vector space of these 7,379 vectors, while continuing to focus on the cited dimension. Closeness centrality cannot be computed in the vector space since the network is not fully connected. Betweenness centrality and degree correlate at $r = 0.69$ ($p < 0.01$). Table 4 provides the top ten journals on these two indicators.

| | *Degree* | | *Betweenness* |
|---|---|---|---|
| *Science* | 0.979534 | *Science* | 0.2860 |
| *Nature* | 0.958254 | *Nature* | 0.2106 |
| *Sci Am* | 0.950935 | *Sci Am* | 0.1946 |
| *J Am Stat Assoc* | 0.942667 | *J Am Stat Assoc* | 0.1785 |
| *Ann NY Acad Sci* | 0.935484 | *Brit Med J* | 0.1471 |
| *P Natl Acad Sci USA* | 0.928707 | *Lancet* | 0.1469 |
| *Lancet* | 0.925047 | *Ann NY Acad Sci* | 0.1409 |
| *Biometrika* | 0.921523 | *Am Econ Rev* | 0.1366 |
| *New Engl J Med* | 0.910952 | *P Natl Acad Sci USA* | 0.1363 |
| *JAMA-J Am Med Assoc* | 0.898075 | *Biometrika* | 0.1350 |

**Table 4**: Top-10 journals in the vector space (being-cited direction).

Seven of the ten journals occur on both lists, and the order of the top four is the same. There are important differences from the top-10 lists provided in table 3. However, it is no longer clear what we are measuring. Both measures correlate, for example, at the level of $r = 0.47$ ($p < 0.01$) with the impact factor, but in themselves they don't have a clear interpretation other than the fact that *Science* and *Nature* have the highest centrality at the global level, no matter how one measures the indicator.

## 6. The local citation impact environments

### 6.1. *Social Networks* as an example

Let us return to our example of the journal *Social Networks* for a more precise understanding of what centrality measures may mean in local citation environments. *Social Networks* is included in the *Social Sciences Citation Index*, but it relates also to journals which are included in the *Science Citation Index*. In the combined set, *Social Networks* is cited by 54 journals (as against 40 in the *Social Sciences Citation Index*). Figure 3 provides the visualization of these journals with the cosine as the similarity measure. The vertical and horizontal axes of the vertices are proportional to the citation impact in this environment with and without within-journal citations, respectively.

Eleven journals are grouped in the bottom right corner because they are isolates in this context. *Social Networks,* and to a lesser extent the *Journal of Mathematical Sociology,* are central in relating major clusters such as two groups of social-science journals (sociology and management science), a physics group, and a group of computer-science journals and statistics. However, the contribution of the two centrally positioned journals to the citation impact in this network is extremely small: only 0.41 % for *Social Networks* and 1.07% for the *Journal of Mathematical Sociology.*

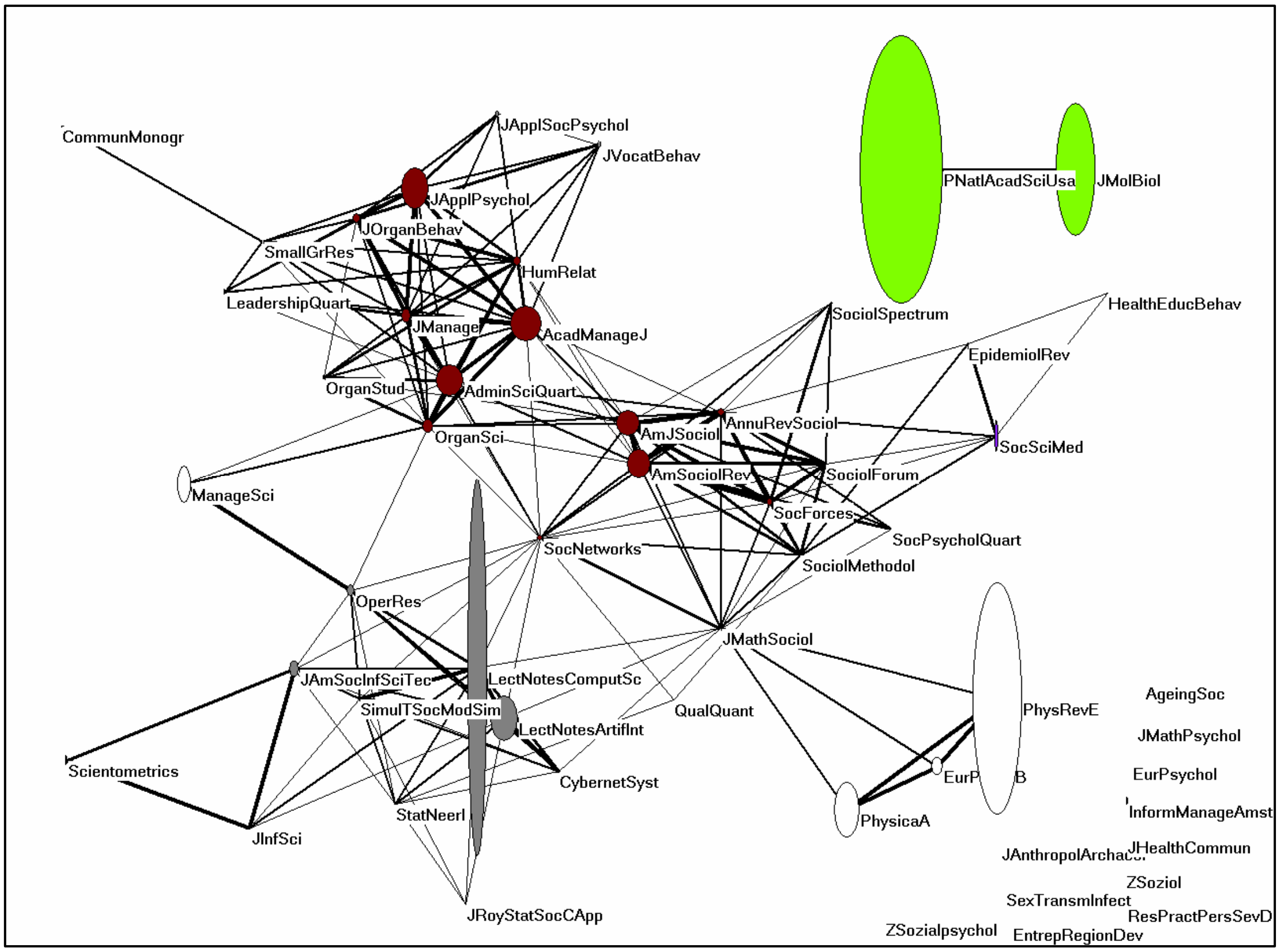

**Figure 3**: Citation impacts of fifty-four journals which cited *Social Networks* more than once in 2004 ($N = 7{,}379$; cosine $\geq 0.2$).

Visual inspection of Figure 3 suggests that these two journals (*Social Networks* and the *Journal of Mathematical Sociology*) are central in relating the various clusters. Using betweenness as a measure, Pajek enables us to draw the vectors for the various measures of centrality and to display the vertices in terms of the values of these vectors. In Figure 1 above, "betweenness centrality" was thus used as the indicator in this same environment.

**Correlations**

| | | Degree | Between-ness | Closeness | Local impact |
|---|---|---|---|---|---|
| Degree | Pearson Correlation | 1 | .724(**) | .877(**) | -.009 |
| | Sig. (2-tailed) | | .000 | .000 | .949 |
| | N | 54 | 54 | 54 | 54 |
| Betweenness | Pearson Correlation | .724(**) | 1 | .542(**) | -.035 |
| | Sig. (2-tailed) | .000 | | .000 | .801 |
| | N | 54 | 54 | 54 | 54 |
| Closeness | Pearson Correlation | .877(**) | .542(**) | 1 | -.001 |
| | Sig. (2-tailed) | .000 | .000 | | .991 |
| | N | 54 | 54 | 54 | 54 |
| Local impact | Pearson Correlation | -.009 | -.035 | -.001 | 1 |
| | Sig. (2-tailed) | .949 | .801 | .991 | |
| | N | 54 | 54 | 54 | 54 |

** Correlation is significant at the 0.01 level (2-tailed).

**Table 5**: Correlations among the different centrality measures and the local impact of 54 journals citing *Social Networks* in 2004.

Table 5 shows that there is no relation between the different centrality measures and the local impact. The Pearson correlation coefficients are negative and not significant. Among the different centrality measures all relations are positive and significant despite the smaller set ($N = 54$).

| | *Degree* | | *Between-ness* | | *Closeness* |
|---|---|---|---|---|---|
| *Soc Networks* | 0.3207547 | *Soc Networks* | 0.152639 | *Soc Networks* | 0.467237 |
| *Acad Manage J* | 0.2830189 | *J Math Sociol* | 0.111096 | *Annu Rev Sociol* | 0.427752 |
| *J Math Sociol* | 0.2641509 | *Annu Rev Sociol* | 0.064244 | *Am J Sociol* | 0.421811 |
| *Admin Sci Quart* | 0.2641509 | *Organ Sci* | 0.055370 | *Organ Sci* | 0.416033 |
| *Am J Sociol* | 0.2641509 | *Acad Manage J* | 0.054492 | *Am Sociol Rev* | 0.416033 |
| *Annu Rev Sociol* | 0.2641509 | *Am J Sociol* | 0.040420 | *Acad Manage J* | 0.410410 |
| *Organ Sci* | 0.2641509 | *Admin Sci Quart* | 0.039110 | *Admin Sci Quart* | 0.404938 |
| *Am Sociol Rev* | 0.2452830 | *J Am Soc Inf Sci Tec* | 0.031328 | *J Math Sociol* | 0.394420 |
| *Hum Relat* | 0.2452830 | *Am Sociol Rev* | 0.030103 | *J Manage* | 0.370370 |
| *J Manage* | 0.2264151 | *Oper Res* | 0.029685 | *Sociol Methodol* | 0.370370 |

**Table 6**: Top-10 journals on the three centrality measures among the 54 journals citing *Social Networks* in 2004.

Table 6 provides the top-10 journals using the three centrality measures in this environment. The various centrality measures are not unrelated, both numerically and for conceptual reasons. A star, for example, with a high degree centrality can be expected to have also a high betweenness centrality, because if the star were removed from the center, the configuration would fall apart. Closeness is defined at the level of the set, and thus less related to betweenness ($r = 0.542$; $p < 0.01$). The main difference which is visible in Table 6 is the skewness of the distribution for "betweenness centrality" when compared with the other two measures. Only a few journals have a high "betweenness centrality."

Figures 4 and 5 below show the degree centrality and the closeness centrality for this same set, respectively.

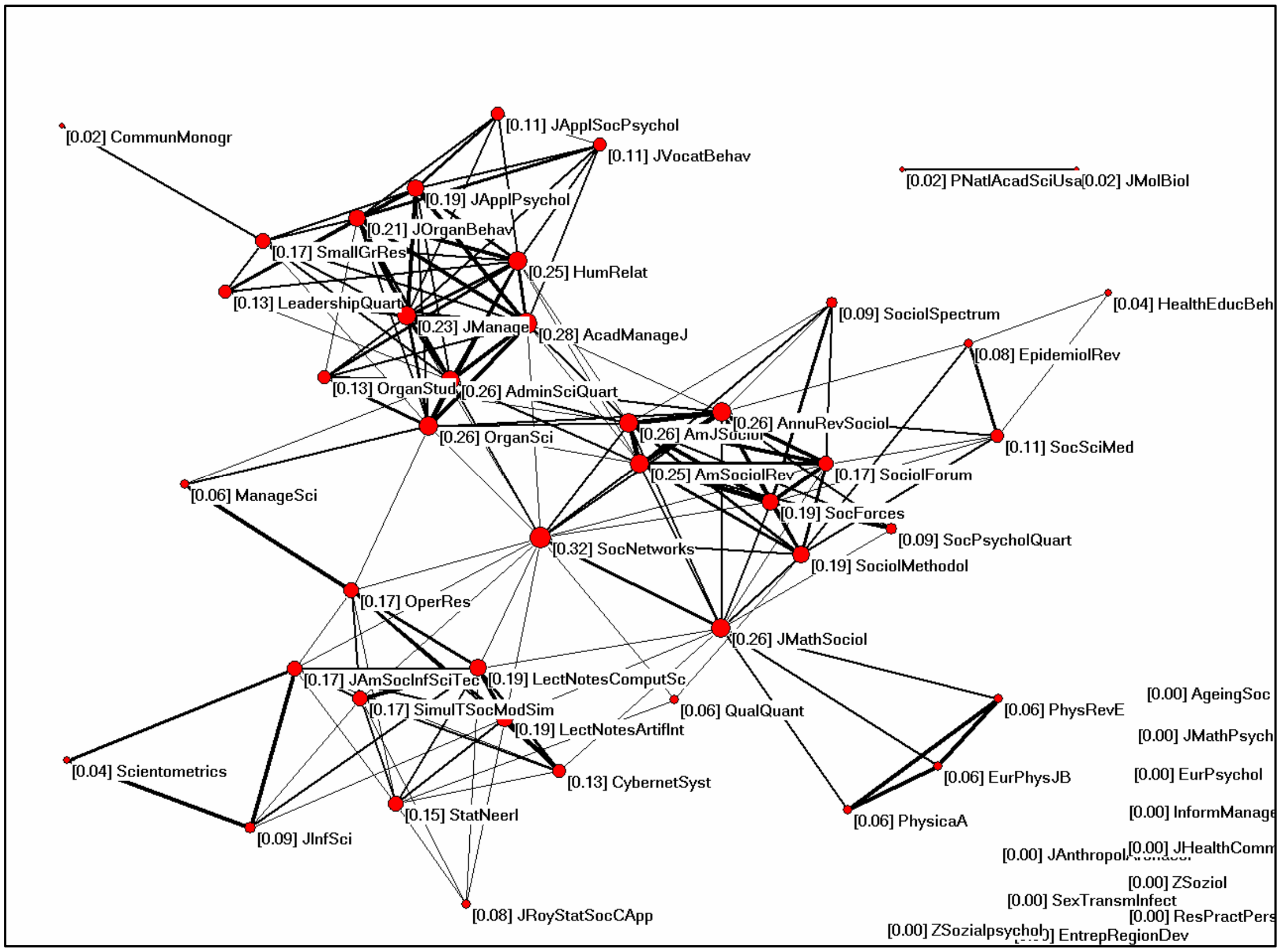

**Figure 4**: Normalized degree centrality of 54 journals in the citation impact environment of *Social Networks* in 2004 (cosine ≥ 0.2).

While *Social Networks* still has the highest value (0.32) in terms of degree centrality in this set—which is probably a consequence of choosing this journal as the seed journal for the construction of the network—the *Journal of Mathematical Sociology* (0.26) is surpassed on this indicator by the *Academy of Management Journal,* which has a value of 0.28 on the indicator. All journals which interface a specific cluster with *Social Networks* and with the other clusters score high on this indicator because these journals construct the coherence of the network.

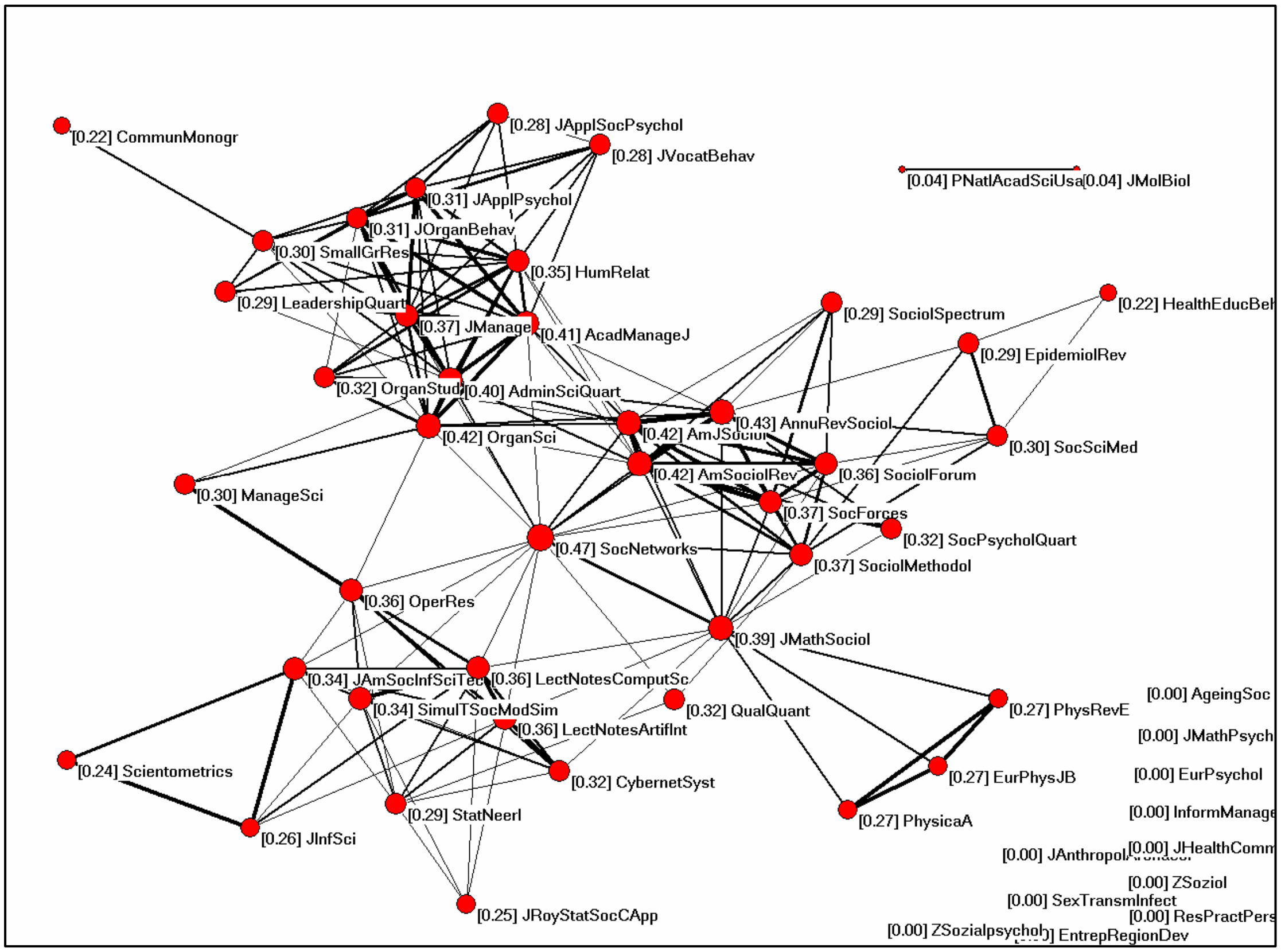

**Figure 5**: Closeness centrality of 54 journals in the citation impact environment of *Social Networks* in 2004 (cosine ≥ 0.2).

"Closeness centrality" (Figure 5) shows also a highest value for *Social Networks*, but various other journals in the set show similarly high values. In summary, *Social Network* is the most central member of this set on all centrality measures. The *Journal of Mathematical Sociology* was second on betweenness and degree, but much less pronounced in the case of "closeness" and "degree centrality." The specific position of the two journals *between* the other clusters visible on the maps is indicated by the measure of "betweenness centrality."

## 6.2 Multi- and interdisciplinarity

While "betweenness centrality" measures the interrelationships among vertices—albeit after normalization because of the possible size effects—"closeness centrality" can be expected to provide us with a more global measure of relationships between groups of vertices. "Closeness" measures relatedness to the set of other vertices. Unlike "interdisciplinarity", "multidisciplinarity" can perhaps be associated with this measurement of different bodies of literature (Klein, 1990). However, as we have seen in Figure 5, the closeness centrality in the cited dimension did not provide us with a strong indicator. In a local environment, "closeness" seems an insufficient discriminator.

The notion of reading different bodies of literature together raises the question of whether perhaps "multidisciplinarity" should be studied using "closeness" as an indicator in the *citing* dimension. Isn't it writing and reading from a variety of sources that makes a journal multidisciplinary? Figure 6 shows the map with closeness centralities of the same 54 journals based on the cosines among the citing patterns.

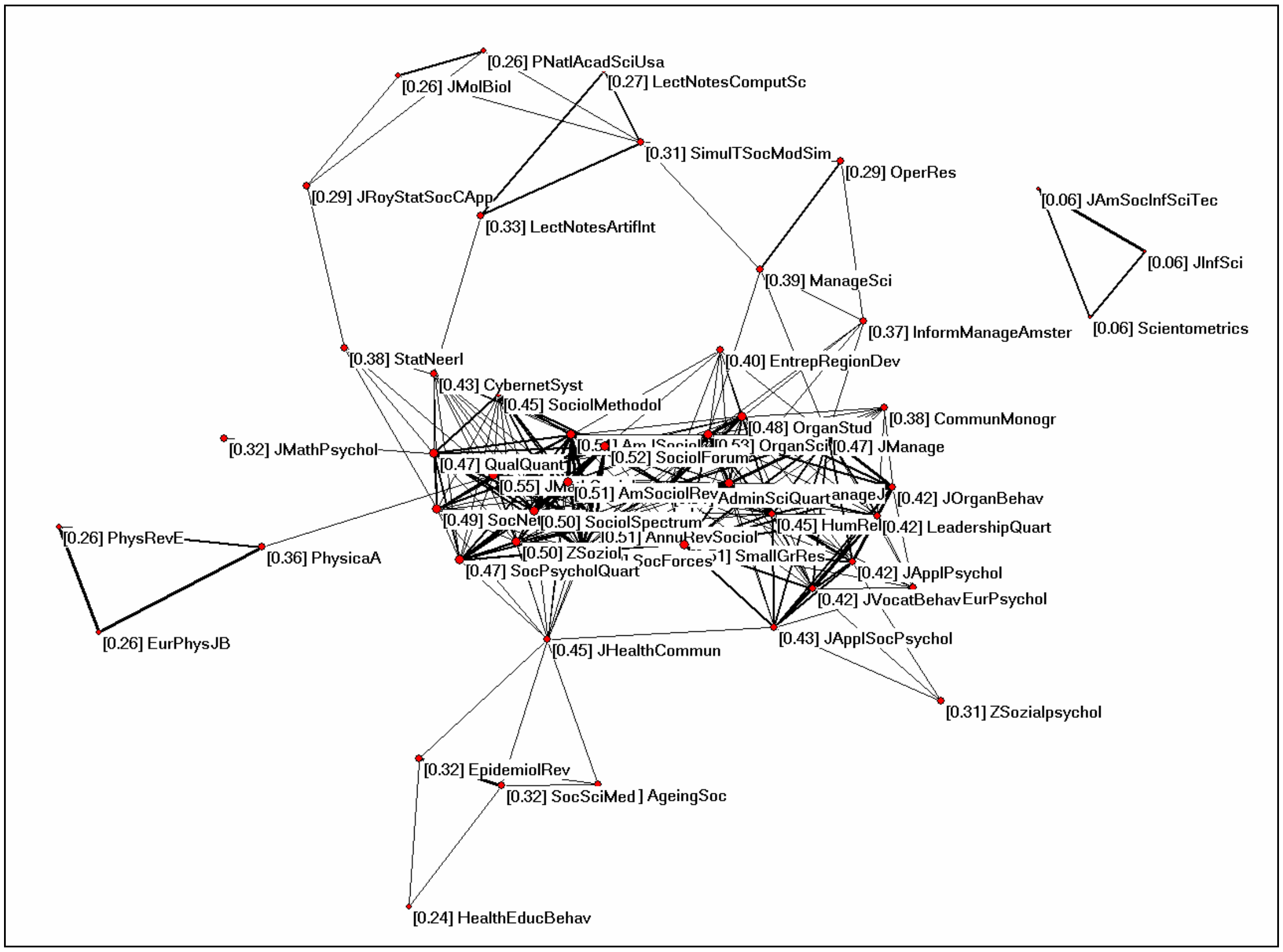

**Figure 6**: Closeness centrality in the citing patterns among the 54 journals cited by *Social Networks* in 2004 (cosine ≥ 0.2).

In this vector space, *Social Networks* is deeply embedded in a cluster of sociology journals which publish methodological contributions. The highest value for closeness centrality among these journals (0.55) is for the *Journal of Mathematical Sociology*. Other journals follow with slightly lower values, among them *Social Networks* with 0.49. Closeness centrality seems to be determined more by the embeddedness of the journal under study within a cluster than to be interpretable in terms of the "multidisciplinarity" of this journal.

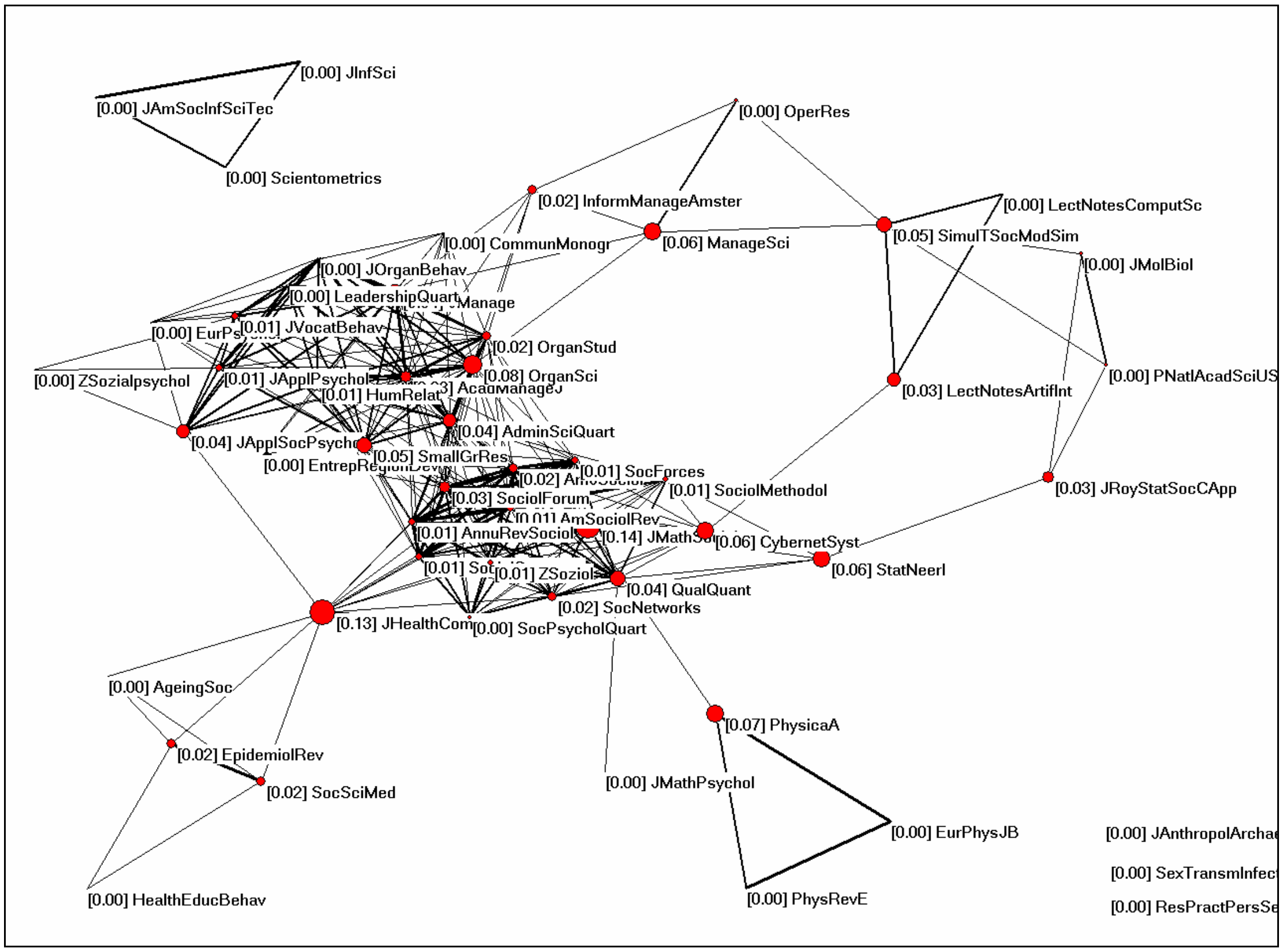

**Figure 7**: Betweenness centrality in the citing patterns among the 54 journals citing *Social Networks* in 2004.

How different is the picture for "betweenness centrality"? As before, the highest values in Figure 7 are attributed to the journals which function as hubs in the network. For example, the *Journal of Health Communication* in the lower left quadrant relates a cluster of journals about social medicine and epidemiology to the core clusters of the field and has the relatively high value of 13% betweenness centrality. The highest value, however, is again for the *Journal of Mathematical Sociology* (14%). *Social Networks* follows in the citing dimension with only 2% betweenness centrality. The relatively high value of 7% for *Physica A* relating a cluster of physics journals to the network analysis journals is noteworthy.

In summary Figures 6 and 7 confirm that "betweenness centrality" can be considered as an indicator of "interdisciplinarity," while the hypothesis that one can use "closeness centrality" as an indicator of "multidisciplinarity" has to be discarded. The journals with high values on "betweenness centrality" in the citing dimensions are those in which authors draw on different litteratures for their citations. These combinations can find their origin in the substantive or the mathematical character of the communications. Journals that are more deeply embedded in a disciplinary cluster tend to show very low values on the betweenness indicator.

| *Centrality* | *Cited* | $(H_{max} - H)/H_{max}$ | *Citing* | $(H_{max} - H)/H_{max}$ |
|---|---|---|---|---|
| Degree | 5.18 bits | 10.0% | 5.29 bits | 8.0% |
| **Betweenness** | **4.11 bits** | **28.6%** | **4.65 bits** | **19.1%** |
| Closeness | 5.36 bits | 6.9% | 5.59 bits | 2.9% |

**Table 7**: Uncertainty in the distribution of the three centrality measures and the reduction of maximum entropy.

In Table 7, entropy statistics is used to show the different shapes of the distributions in the three measures of centrality (Theil, 1972; Leydesdorff, 1995). The reduction of uncertainty in the distribution (when compared with the maximal uncertainty) is 28.6% in the cited dimension and 19.1% in the citing dimension for the "betweenness centrality" measure, while it is only 6.9% and 2.9%, respectively, in the case of "closeness centrality." Thus, the measure of "betweenness centrality" is specific and therefore discriminatory for specific journals, much more than the measure of "closeness centrality." The specificity of "betweenness centrality" is a consequence of its local character and the normalization implied by using the vector space. Table 8 shows the same information in terms of the skewness and kurtosis of these distributions.[2]

| | N | Skewness | | Kurtosis | |
|---|---|---|---|---|---|
| | Statistic | Statistic | Std. Error | Statistic | Std. Error |
| Degree | 54 | .301 | .325 | -1.131 | .639 |
| **Betweenness** | **54** | **3.073** | **.325** | **11.255** | **.639** |
| Closeness | 54 | -.808 | .325 | -.765 | .639 |

**Table 8**: Descriptive statistics of the skewness and kurtosis in the distributions of the three centrality measures in the vector space.

### 6.3 Larger and smaller sets: domain dependency

Of the 54 journals citing *Social Networks* in 2004, 40 are included in the *Social Sciences Citation Index*. Figure 8 shows that the major interface in this citation impact environment of *Social Networks* is between sociological journals in the top part of the screen and organization & management science in the lower half. At this interface, however, *Social Networks* shares its interdisciplinary position with other journals which maintain this relationship in their aggregated being-cited patterns. It is surpassed on the betweenness indicator by the *Annual Review of Sociology* and the *Academy of*

---

[2] *Skewness* is a measure of the asymmetry of a distribution. The normal distribution is symmetric and has a skewness value of zero. A distribution with a significant positive skewness has a long right tail. A distribution with a significant negative skewness has a long left tail. As a rough guide, a skewness value more than twice its standard error is taken to indicate a departure from symmetry. *Kurtosis* is a measure of the extent to which observations cluster around a central point. For a normal distribution, the value of the kurtosis statistic is 0. Positive kurtosis indicates that the observations cluster more and have longer tails than those in the normal distribution and negative kurtosis indicates the observations cluster less and have shorter tails. (Source: SPSS 13.0, Helpfile.)

*Management Journal.* The reason for this lower value on betweenness is the absence of a statistics and computer science cluster in this database. Thus, the interdisciplinarity of *Social Networks* is reduced to 5% in this citation environment.

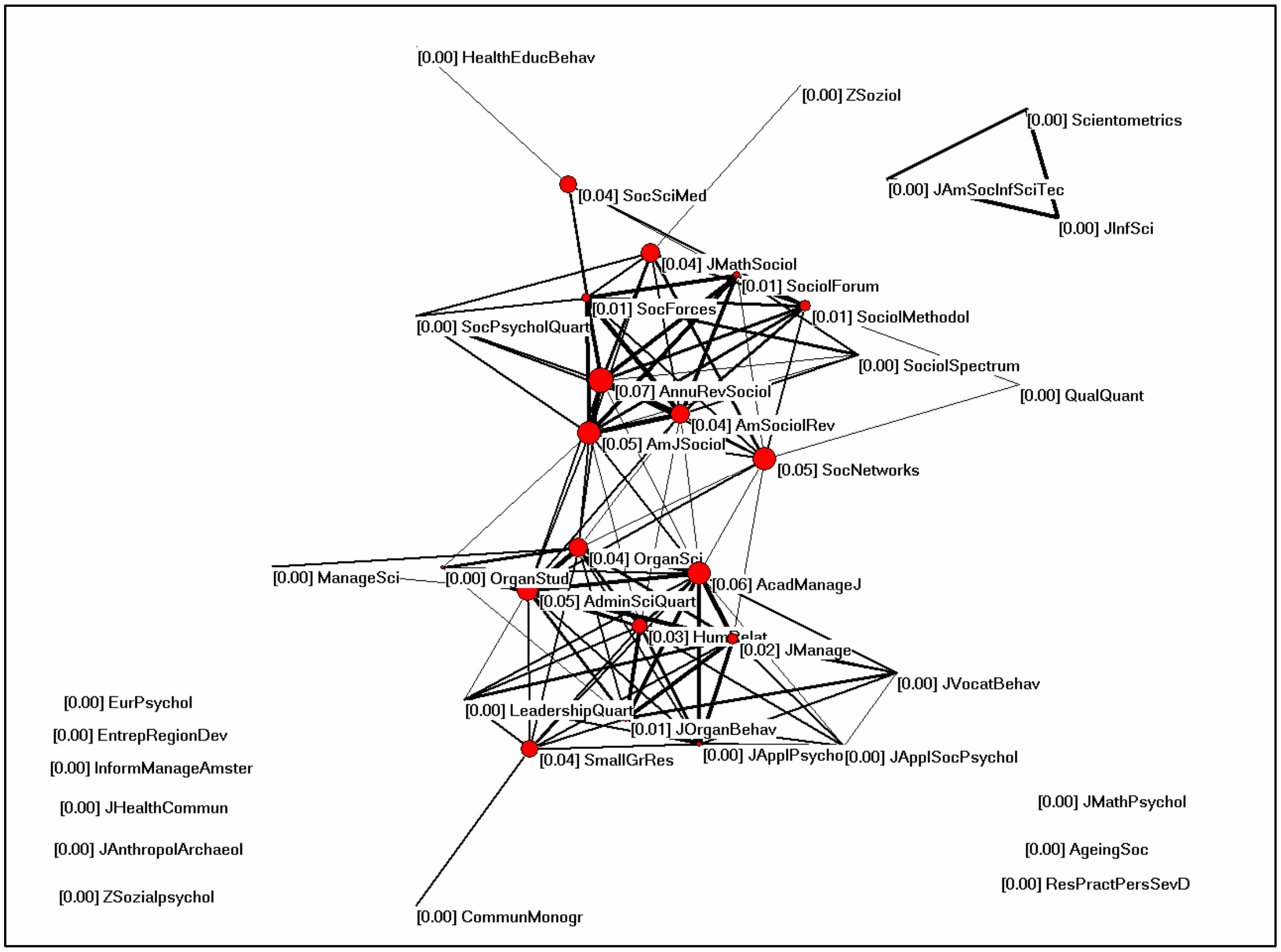

**Figure 8**: Betweenness Centrality among the 40 journals in the citation environment of *Social Networks* using exclusively the ***Social Sciences Citation Index*** as a database (cosine $\geq 0.2$).

In other words, the interdisciplinarity of a journal is relative to the set of journals used for the assessment. Like citation impact, the indicator can only be properly defined with reference to a given citation environment. In the environment of each journal or group of journals, the specific function of "interdisciplinarity" is expected to be different.

In summary, the findings are: 1) betweenness centrality is extremely effective as a measure of "interdisciplinarity"; and 2) closeness centrality fails as a measure of "multidisciplinarity." Betweenness centrality as a measure of "interdisciplinarity" both improves and changes as it is applied to sets that are more and more coherently defined. Closeness centrality, however, fails at all levels as a measure of "multidisciplinarity."

## 7. Further tests and applications

In this section, the betweenness centrality measure in the vector space is applied to three more cases. First, *Scientometrics* is used as a seed journal in order to demonstrate that the high value for betweenness centrality in the previous case (of *Social Networks*) was not a consequence of choosing the journal as a seed journal. One can expect the seed journal to have a high degree centrality—although not necessarily the highest—in its own citation environment, but the normalization using the cosine precisely corrects for the effect of degree centrality on betweenness centrality in the vector space.

In the other two cases, I return to policy applications by focusing on biotechnology and nanotechnology. In the case of biotechnology, a cluster of journals has been formed during the past two decades. Within this cluster some journals on the technological side are more "interdisciplinary" than others in terms of the proposed measure. In the case of nanotechnology, the set has not yet crystallized. However, the betweenness measure highlights the journals at the interface which are most central to the emerging field of nanoscience and nanotechnology.

## 7.1 *Scientometrics*

Let us consider *Scientometrics* as a seed journal and generate the set of 54 journals which cited this journal more than once in the aggregate during 2004. Using the proposed indicator, Figure 9 was generated among the 30 of these 54 journals which form a graph together above the level of cosine ≥ 0.2. The fact that the other 24 journals are not related to the core set at this level illustrates the incidental citation relations which *Scientometrics* as a journal maintains with various disciplines. Two of these groups of journals are visible in Figure 9 because they are related among themselves.

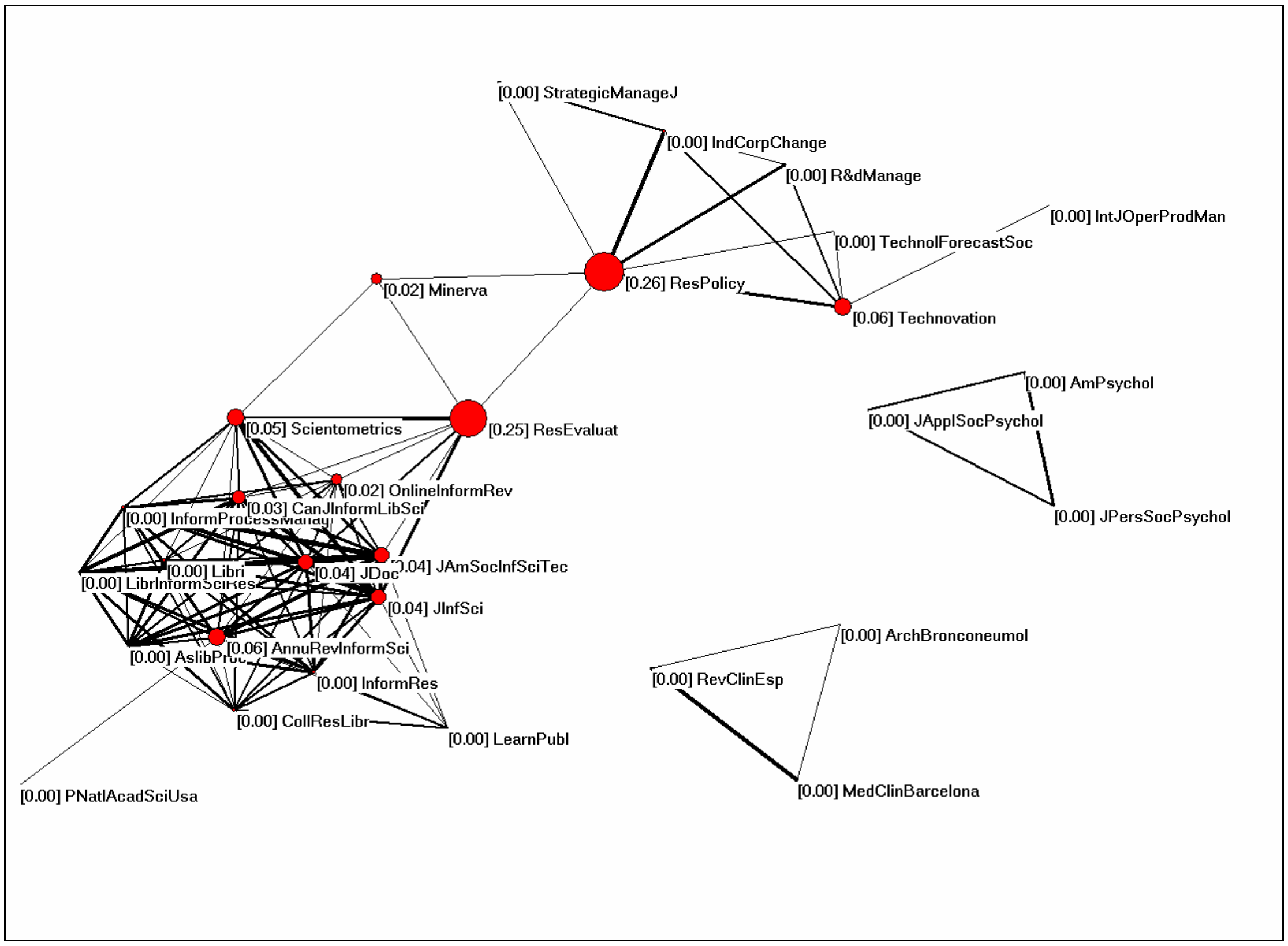

**Figure 9**: Betweenness centrality in the citation environment of *Scientometrics* in 2004 (cosine ≥ 0.2).

Although *Scientometrics* scores with a betweenness centrality of 5% above the other information science journals, *Research Policy* (26%) and *Research Evaluation* (25%) are the two interdisciplinary journals in this environment. The other journals are embedded in disciplinary clusters.

### **7.2** *Biotechnology and Bioengineering*

For the field of *Biotechnology and Bioengineering* I used the core journal of this field with this same name. This journals was cited in 2004 by 633 journals, of which 16 contributed to its citation environment to the extent of more than one percent of its total citation rate (11,652). Figure 10 shows the local citation impacts of these journals (He & Pao, 1986; Leydesdorff & Cozzens, 1993). The journal is central to a cluster of biotechnology journals. Its ISI-impact factor is 2.216.

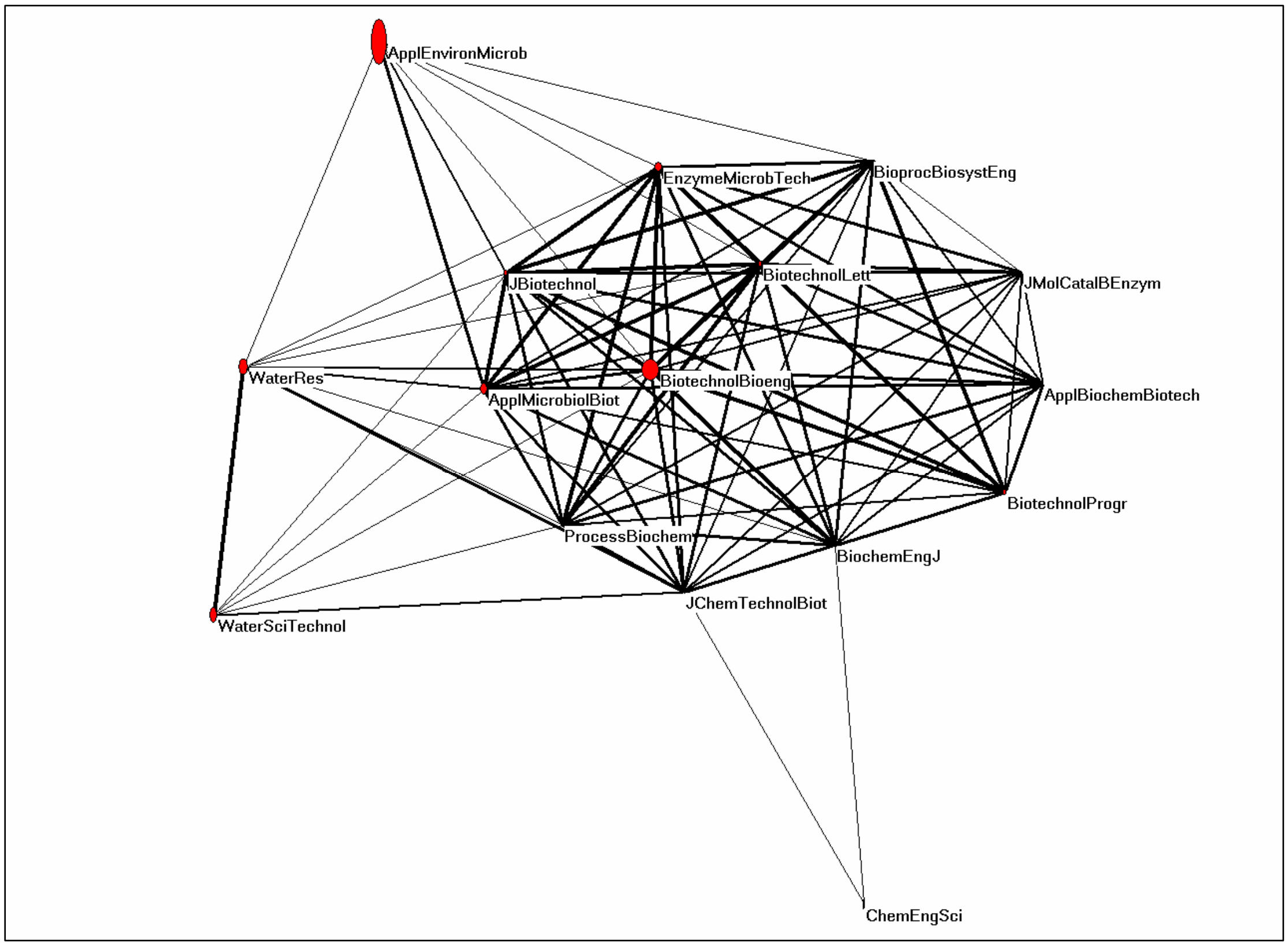

**Figure 10**: Citation impact environment of *Biotechnology and Bioengineering* in 2004 (threshold 1%; cosine ≥ 0.2)

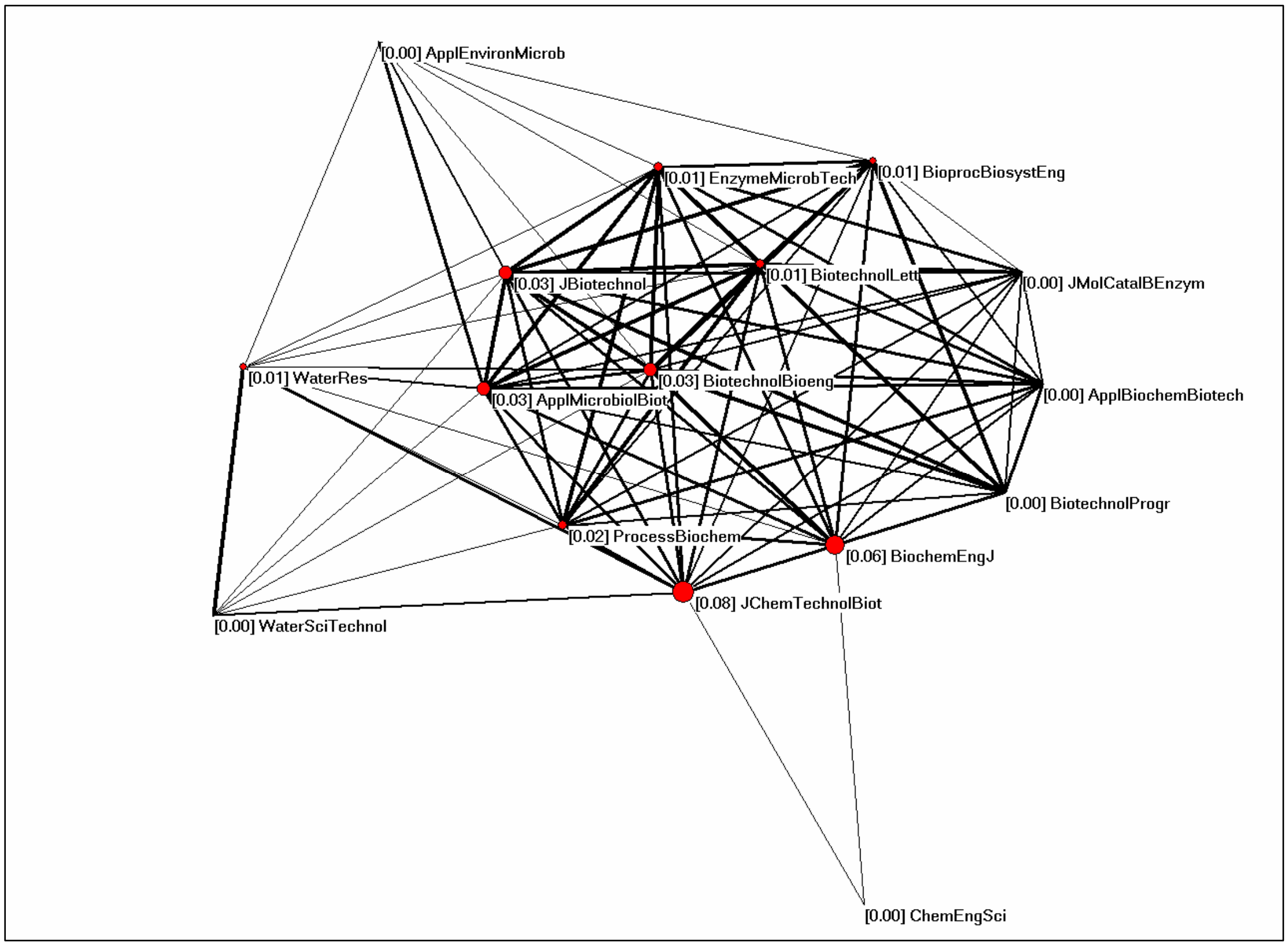

**Figure 11**: Betweenness centrality of 16 journals citing *Biotechnology and Bioengineering* in 2004 (threshold 1%, cosine ≥ 0.2).

Figure 11 shows that the betweenness centrality is not distributed equally among the cluster of journals which form the central group. The *Journal of Chemical Technology and Biotechnology* has a betweenness centrality of 8%, while *Biotechnology and Bioengineering* has a position on the geodesics among the other journals in only 3% of the possible ones. The *Biochemical Engineering Journal* is second on this indicator with a value of 6%. Thus, the focus of the interdisciplinary development of biotechnology is to be found on the engineering side of the field.

## 7.3 *Nano Letters*

Three hundred and five journals constitute the citation environment of *Nano Letters,* which can be considered as a leading journal in the field of nanoscience and nanotechnology (Zhou & Leydesdorff, 2006). *Nano Letters* has an impact factor of 8.449. However, Figure 12 shows that the local citation impact among the 17 journals that cite *Nano Letters* to the extent of more than 1% of its total number of citations (7,349) is modest at best. The journal is embedded in an ecology of journals which include major physics and chemistry journals with considerably higher citation rates in this environment.

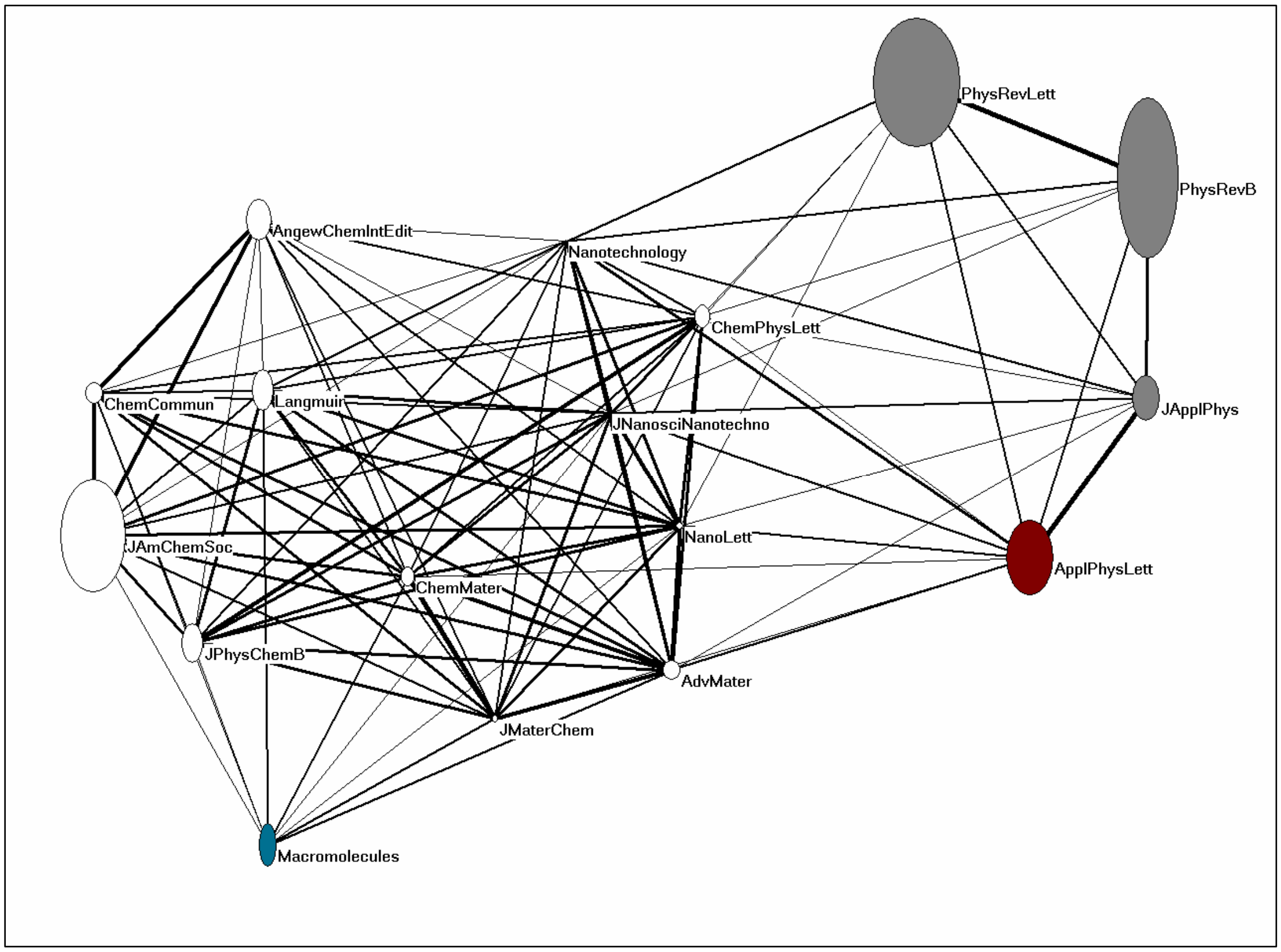

**Figure 12**: Citation impact of 17 journals in the environment of *Nano Letters* 2004 (threshold 1%; cosine ≥ 0.2).

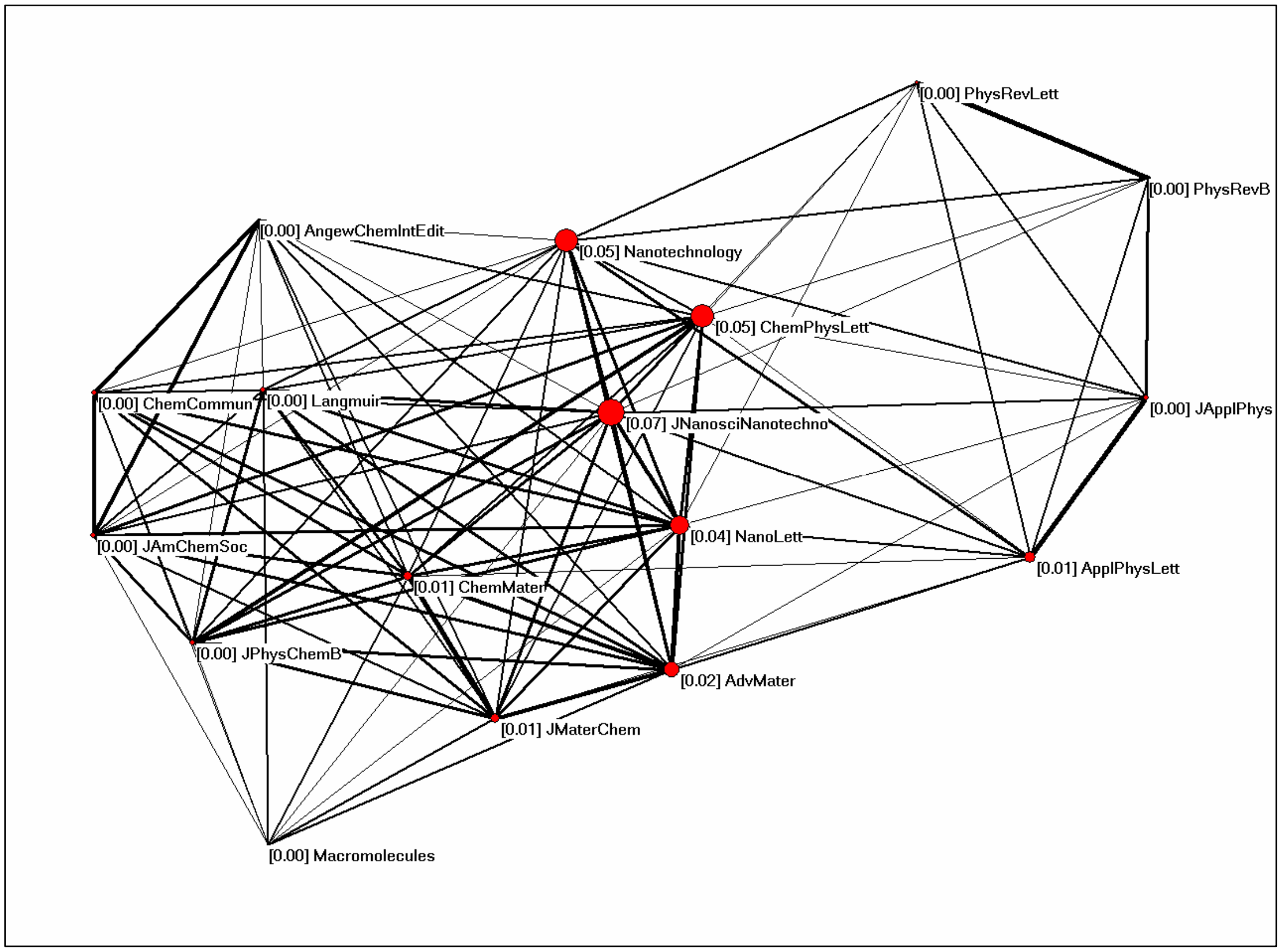

**Figure 13**: Betweenness centrality of 17 journals in the citation impact environment of *Nano Letters* 2004.

The betweenness centrality measure shows that *Nano Letters* belongs to a group of journals with an interdisciplinary position between the surrounding chemics and physics journals. However, in terms of betweenness centrality the *Journal of Nanoscience and Nanotechnology* has a higher value (7%) than *Nano Letters* (4%) despite its much lower impact factor (2.017) and total citation rate (489). Figure 13 shows the interdisciplinary interface of nano-science journals in terms of the sizes of the nodes.

## 8. Conclusions

The finding of betweenness centrality as a possible indicator of interdisciplinarity was originally a serendipitous result of my work on journal mapping. The conclusion that an unambiguous classification of the journal sets in terms of their aggregated citation patterns is impossible because of the multi-dimensionality of the space and the fuzziness of important delineations (Bensman, 2001; Leydesdorff, 2006) led me to consider the input information for drawing maps from the perspective of any of the journals included in the *Science Citation Index* online (at http://www.leydesdorff.net/jcr04) so that the user can draw maps using Pajek or a similar visualization program (Leydesdorff, forthcoming-b). One can use the same files and pictures for mapping the betweenness of the interdisciplinary journals by choosing the option of this indicator for the visualization. In addition to visualizing the nodes in proportion to their respective values on this vector, Pajek provides the quantitative information about the vector values.

I should once more emphasize that the betweenness centrality in the vector space is computed on a matrix completely different from that used for the computation of centrality in the multidimensional data space. The latter measures are computed from the asymmetrical citation matrix among the journals and indicate the degree of centrality of journals in the cited dimension. The measure for "interdisciplinarity" suggested here, however, is based on the symmetrical cosine matrix which was constructed on the basis of the cosines among these vectors. This normalization has the advantage of controling for the absolute size of a journal and its consequently higher probability of degree centrality.

The analogous expectation that "closeness centrality" might provide us with an indicator of "multidisciplinarity" could not be substantiated. In contrast to interdisciplinarity, multidisciplinarity would mean, in my opinion, that a journal prints articles from different disciplinary backgrounds without necessarily integrating them. The *Lecture Notes in Computer Science* are a case in point since this large collection of proceedings is not substantively integrated. The closeness measure, however, remains a global measure which loses its discriminating power in a local context when the specific relations are already tightly knit. At the level of the global set, the problem of the heterogenity among subsets prevails, and therefore conclusions cannot be drawn.

In the specific case of the citation environment of *Social Networks* elaborated in this study, it seemed first in the cited dimension that the choice of the seed journal was decisive for centrality on all three measures. However, in the citing dimension *Social Networks* lost this interdisciplinary position; other journals in the field are primary in the reconstruction of the structure. The restriction to journals included in the *Social Sciences Citation Index* showed that the measure is domain-specific. In further applications, the discriminating power of the betweenness measure could convincingly be shown.

The advantage of the measure is that it can be applied directly to the matrices of cosine values that were brought online at http://www.leydesdorff.net/jcr04. One can import these (ASCII) files into Pajek and choose the betweenness centrality option for the measurement and visualization. The value of the threshold in the cosine (≥ 0.2) is arbitrary, but sufficiently low to show all important connections. (A threshold has to be set because the cosine between citation patterns of locally related journals will almost never be zero, and the algorithm for betweenness centrality first binarizes the matrix.) I intend to bring the matrices based on the *Journal Citation Reports* 2005 shortly online so that one can study interdisciplinarity using this measure for the most recent data.

**Acknowledgments**

I am grateful to Janusz Holyst, Diana Lucio Arias, and Ronald Rousseau for comments and suggestions.